\documentclass{article}

\usepackage{microtype}
\usepackage{graphicx}
\usepackage{subfigure}
\usepackage{booktabs}
\usepackage{hyperref}

\usepackage[accepted]{icml2025}

\usepackage{amsmath}
\usepackage{amssymb}
\usepackage{mathtools}
\usepackage{amsthm}
\usepackage[capitalize,noabbrev]{cleveref}
\usepackage[textsize=tiny]{todonotes}

\theoremstyle{plain}

\theoremstyle{definition}

\theoremstyle{remark}

\icmltitlerunning{HiLWS: Human-in-the-Loop Weak Supervision for Remote Neurological Video Assessment}

\begin{document}

\twocolumn[
\icmltitle{HiLWS: A Human-in-the-Loop Weak Supervision Framework for Curating Clinical and Home Video Data for Neurological Assessment}

\begin{icmlauthorlist}
\icmlauthor{Atefeh Irani}{ubc,tehran}
\icmlauthor{Maryam S. Mirian}{ubc}
\icmlauthor{Alex Lassooij}{ubc}
\icmlauthor{Reshad Hosseini}{tehran}
\icmlauthor{Hadi Moradi}{tehran}
\icmlauthor{Martin J. McKeown}{ubc}
\end{icmlauthorlist}

\icmlaffiliation{ubc}{Pacific Parkinson Research Centre, University of British Columbia, Vancouver, BC, Canada}
\icmlaffiliation{tehran}{School of Electrical and Computer Engineering, University of Tehran, Tehran, Iran}

\icmlcorrespondingauthor{Martin J. McKeown}{martin.mckeown@ubc.ca}

\icmlkeywords{Machine Learning, ICML}
\vskip 0.3in
]
\printAffiliationsAndNotice{} 
\begin{abstract}
Video-based assessment of motor symptoms in conditions such as Parkinson’s disease (PD) offers a scalable alternative to in-clinic evaluations, but home-recorded videos introduce significant challenges, including visual degradation, inconsistent task execution, annotation noise, and domain shifts. We present \textbf{HiLWS}, a cascaded human-in-the-loop weak supervision framework for curating and annotating hand motor task videos from both clinical and home settings. Unlike conventional single-stage weak supervision methods, \textbf{HiLWS} employs a novel cascaded approach, first applies weak supervision to aggregate expert-provided annotations into probabilistic labels, which are then used to train machine learning models. Model predictions, combined with expert input, are subsequently refined through a second stage of weak supervision. The complete pipeline includes quality filtering, optimized pose estimation, and task-specific segment extraction, complemented by context-sensitive evaluation metrics that assess both visual fidelity and clinical relevance by prioritizing ambiguous cases for expert review. Our findings reveal key failure modes in home recorded data and emphasize the
importance of context-sensitive curation strategies for robust medical video analysis.
\end{abstract}

\section{Introduction}

Remote health monitoring is increasingly employed to improve healthcare accessibility, facilitate early detection, and enable continuous tracking of chronic conditions. Within this paradigm, video-based assessments provide a non-invasive and scalable means of evaluating motor and cognitive function in real-world settings \cite{deng2024interpretable}. These assessments often rely on structured tasks, such as finger tapping, opening-closing of the hand, and facial expressions, that serve as validated proxies for neurological evaluation, particularly in conditions like Parkinson's Disease (PD).

PD is characterized by progressive motor impairment, and clinical evaluation commonly involves tasks from the Movement Disorder Society-sponsored Unified Parkinson’s Disease Rating Scale (MDS-UPDRS) \cite{goetz2008movement}. While clinic-based assessments provide structured and standardized observations, deploying video-based evaluations in home settings introduces challenges related to task adherence, data quality, and label reliability. These problems are further compounded by contextual domain shifts between structured clinical and uncontrolled home recordings.

Recent advances in pose estimation (e.g., MediaPipe \cite{lugaresi2019mediapipe}, OpenPose \cite{qiao2017real}) and federated learning \cite{ali2022federated} have enabled privacy-preserving and distributed model development. Landmark-based representations also reduce storage and computational costs. However, real-world deployments remain hindered by annotation noise, inconsistent video quality, and demographic or environmental biases that affect model generalization \cite{rahmani2021machine}.

We introduce \textbf{HiLWS}, a cascaded human-in-the-loop weak supervision framework for curating video recordings of standardized hand motor tasks captured in both clinical and home environments. Unlike prior approaches that rely on fully labeled, single-domain datasets, \textbf{HiLWS} is explicitly designed to handle heterogeneous data quality and annotation sparsity across domains. Our modular pipeline first aggregates noisy expert labels using weak supervision and then leverages these aggregated labels to train task-specific models whose output inform a second-stage refinement process. Using structured clinical recordings as reference, we systematically characterize failure modes in home-recorded videos and demonstrate that conventional labeling strategies are insufficient in this high-stakes medical context. The pipeline incorporates adaptive quality filtering, optimized pose estimation, task-specific segmentation, and context-sensitive metrics to guide expert review toward ambiguous cases. Our findings highlight the need for domain-aware evaluation and selective expert adjudication to ensure both model reliability and clinical safety in real-world deployment.

Our contributions are threefold: (1) a modular video curation pipeline that integrates adaptive quality filtering, pose estimation tuning, and task-aware segment selection; (2) the \textbf{HiLWS} labeling system, which combines expert and machine-generated annotations into a probabilistic framework; and (3) a suite of evaluation metrics that link technical artifacts to clinically meaningful errors. Through systematic degradation and sensitivity analyses, we identify key quality thresholds—such as minimum frame rate and hand-to-frame ratio—below which model performance degrades significantly. We validate the framework on a heterogeneous dataset of over 2,000 videos from both clinical and home environments, demonstrating which curation strategies are broadly generalizable and which require context-specific adaptation for reliable deployment.

This study proposes a generalizable methodology for the robust curation of noisy medical video data and offers broader insights into how data-centric strategies must adapt to diverse real-world acquisition conditions—aligning with the DataWorld workshop’s mission to bridge domain-specific approaches with cross-context curation practices.

\section{Related Work}

\textbf{Video-Based Assessment of Motor Tasks.}
Video-based analysis has become a vital tool for assessing motor symptoms in neurological disorders, particularly PD. Systems like VisionMD \cite{acevedo2025visionmd} provide open-source, automated scoring by computing kinematic features extracted from pose estimation frameworks, enabling objective monitoring over time. Spatio-temporal graph convolutional networks have achieved high accuracy in classifying motor tasks from video, using joint landmark sequences aligned with MDS-UPDRS \cite{tian2024cross}. Recent approaches leverage markerless pose estimation models such as MediaPipe to extract body landmarks from standard video, which are then used to compute movement-specific features including tremor amplitude, arm swing asymmetry, and gait cadence \cite{deng2024interpretable}. These features serve as input to machine learning models—ranging from interpretable tree-based classifiers to deep neural networks—to assess symptom severity, classify task types, or replicate clinician ratings with high fidelity \cite{sarapata2023video,mifsud2024detecting}. Clinically, these systems facilitate scalable, objective, and frequent symptom tracking in home settings, supporting early diagnosis and longitudinal monitoring in telehealth environments. These techniques offer scalable, repeatable, and hardware-light alternatives to traditional clinical assessment.

\textbf{Video Quality in Medical AI.}
Video quality substantially influences the reliability of automated assessments. Existing evaluation strategies can be categorized into perceptual quality-based and task-based methods \cite{kolarik2023explainability}. While perceptual metrics assess visual clarity, task-based approaches emphasize how degradation (e.g., blur, occlusion, low resolution) affects downstream diagnostic accuracy \cite{ye2022effects}. In medical AI, where subtle motion cues inform clinical decision-making, quality distortions pose a critical risk. However, standardized clinical-grade video quality assessment protocols remain underdeveloped, particularly for unconstrained home environments. Recent work underscores the need for application-specific thresholds that account for domain-specific artifacts and context \cite{sibley2021video}.

\textbf{Label Denoising in Medical Applications.}
Medical datasets often suffer from noisy annotations due to inter-rater variability, ambiguous cases, and limited expert availability. Weak supervision frameworks such as Snorkel \cite{ratner2020snorkel} address this by aggregating noisy labels into probabilistic consensus scores. Recent innovations combine contrastive learning and mixup attention mechanisms to suppress noise during training and improve generalization under label uncertainty \cite{gao2024suppressing}. Although these approaches are well-established in static imaging domains, their integration into temporal video tasks for clinical use remains an emerging frontier. Leveraging these denoising techniques in conjunction with expert-in-the-loop adjudication presents a promising direction for improving label fidelity in video-based remote assessments.

\section{Dataset Description}
A total of 2,158 videos were collected from 88 individuals in home environments and 227 individuals in clinical settings. Each subject performed seven standardized hand motor tasks. In all tasks, participants were asked to perform repetitive actions, with the primary objective of evaluating the severity of the disease based on the variability in the execution of the task. Demographic information for clinical and home cohorts is provided in Appendix Table ~\ref{tab:demography} and Table ~\ref{tab:pd_unique_demo}, respectively.

The primary data processing pipeline consists of: landmark extraction, event detection, feature extraction, and medical diagnosis. The severity of the disease is rated on a scale of 0 to 4 for each task, with 4 indicating the most severe impairment.

\begin{figure*}[t]
\centering
\includegraphics[width=\textwidth]{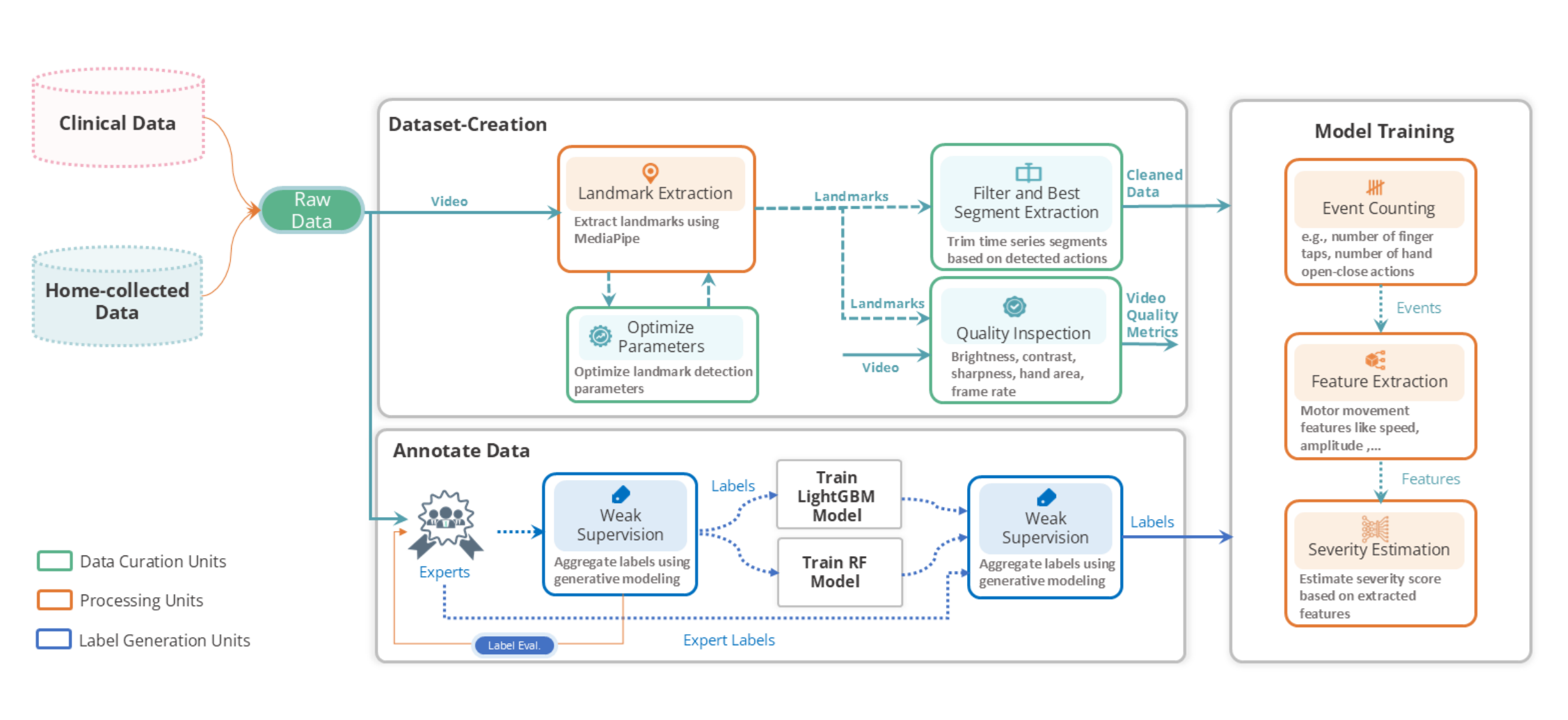}
\caption{Overview of the proposed pipeline for curating and analyzing structured hand motor task videos in remote neurological assessment. The process integrates clinical and home-recorded data, with modules for video quality filtering, landmark-based segment extraction, and probabilistic label fusion via cascaded human-in-the-loop weak supervision (\textbf{HiLWS}). Downstream components include event detection, feature extraction, and symptom severity estimation to support reliable model development from real-world, heterogeneous video sources.}
\label{fig:pipeline}
\end{figure*}

\section{Challenges in Data}
Developing such a system presents several challenges.

\subsection{Visual Variability}

From a visual perspective, factors such as lighting, sharpness, background clutter, resolution, and camera distance varied significantly between clinical and home-recorded videos. While many of these issues can be mitigated by using landmark-based representations, their effectiveness diminishes when visual quality degrades or parameters are not properly fine-tuned. In addition, incorrect or incomplete task execution was frequently observed in home settings, highlighting the need for real-time feedback mechanisms to guide participants during recording.

\subsection{Annotation Noise}

Annotation noise, often resulting from inter-rater disagreement in medical assessments, can substantially undermine label reliability \cite{svingos2023inter}. For instance, incorrectly labeling a home video of a healthy individual as displaying bradykinesia may lead to false positives in remote monitoring systems. In the dataset, approximately 60\% of healthy control (HC) individuals were assigned a non-zero symptom severity score in at least one task. This raises the challenge of treating such samples as zero-label instances without careful reannotation. Simultaneously, Level 0 (no symptom) classifications demand higher annotation specificity and granularity. 

Additionally, the data distribution is heavily imbalanced: only about 29\% of samples belong to the ``no symptom'' class, despite this being the most common presentation in real-world deployments. This skew can bias models toward overestimating symptom severity. Minimizing false positives in the zero-class is thus critical for safe deployment in unsupervised environments. Given the high cost of collecting medical video, model performance becomes tightly coupled to both label accuracy and distributional alignment.

\section{Curation Strategies}

Effective curation is crucial for reliable analysis of medical videos, Figure~\ref{fig:pipeline} illustrates our modular pipeline.

\subsection{Video Quality Parameters}

Quality metrics were evaluated to characterize video attributes. Brightness was quantified as the mean value in the HSV color space to represent overall lighting conditions \cite{okarma2008colour}. Contrast was derived from the L channel in the LAB color space, providing a measure of visual clarity \cite{wang2023contrast}. Sharpness was evaluated using Laplacian variance, capturing the level of image detail \cite{so2024solar}. Relative hand area quantifies the proportion of the video frame occupied by the detected hand, serving as a proxy for subject framing quality and positioning adequacy. Frame rate, extracted from video metadata, was analyzed to determine its impact on temporal resolution and motion accuracy.

\subsection{Task-Aware Filtering}

To ensure that only task-compliant segments are analyzed, a filtering mechanism was implemented to automatically discard irrelevant or poorly executed portions of the recording. This is important for home-recorded videos, where subjects may deviate from the prescribed protocol. For tapping tasks, temporal consistency and frequency heuristics were used to select optimal segments where activity was sustained. This step improves the quality of extracted features and increases the reliability of medical interpretation by focusing analysis on the most informative motion windows.

\subsection{Cascaded Human-in-the-Loop Weak Supervision (HiLWS)}

To address the challenge of scalable yet reliable annotation in home-recorded and clinical videos, a cascaded human-in-the-loop decision-making process was integrated with weak supervision \cite{ratner2020snorkel}. Cascade here refers to the sequential application of label aggregation.\textbf{HiLWS} operates in two stages: an initial expert-only annotation stage, followed by a second-stage fusion that integrates both expert and machine-generated labels using a probabilistic model. The human-in-the-loop component is driven by uncertainty in the first-stage output and is applied selectively to ambiguous cases identified by weak supervision.

In the first stage, five clinical experts independently annotate a subset of the video dataset. Their aggregated  labels with weak supervision are used to train two machine learning models:
(1) \textbf{LightGBM classifier}, trained on time-domain features such as amplitude and velocity extracted from hand landmark trajectories \cite{islam2023using}.
(2) \textbf{Random Forest classifier}, trained on spatial-temporal and frequency-domain features, including tapping rhythm and joint angle variability, selected for its robustness to noise and heterogeneous input.

In the second stage, annotations from all seven sources—the five experts and two models—are combined into another weak supervision model. 

Let \(x_i\) denote a video instance and \(\mathcal{Y} = \{y_i^{(1)}, \ldots, y_i^{(R)}\}\) represent the noisy labels from \(R\) sources. A label model \(f: x_i \rightarrow \hat{p}(y_i)\) is used to produce soft probabilistic labels over the five severity classes.

To identify uncertain predictions requiring expert adjudication, we compute the entropy of the predicted soft label distribution:
\[
H(\hat{p}(y_i)) = - \sum_c \hat{p}_c(y_i) \log \hat{p}_c(y_i),
\]
where \(\hat{p}_c(y_i)\) is the estimated probability of class \(c\) for sample \(x_i\). A high entropy value indicates uncertainty or disagreement among the sources. Samples with high entropy are returned to experts for review, forming the human-in-the-loop feedback loop that refines the final labels only where necessary.

The final hard label is obtained via weighted majority vote:
\[
\hat{y}_i = \arg\max_{c} \sum_{r \in R_i} w_r \cdot \mathbf{1}(\lambda_r(x_i) = c),
\]
where \(R_i \subseteq \{1, \ldots, 7\}\) is the set of sources that provided a label for \(x_i\), \(\lambda_r\) is the labeling function from source \(r\), and \(w_r\) is the learned confidence weight for each source.

This cascaded strategy addresses common challenges in real-world medical video annotation, including sparse expert coverage and inconsistent labels. By using weak supervision to guide targeted expert intervention, \textbf{HiLWS} balances scalability with label quality, producing annotations that are both efficient and clinically trustworthy.

\section{Evaluation Metrics}
Five metrics are used to evaluate different components of the system, including video quality, landmark detection, event detection, and medical evaluation.

\subsection{Landmark Detection Failure Rate (LDFR)}
Failure is defined as the inability to detect a hand or the detection of an implausible number of hands. The failure rate is calculated as:
\[
\text{Failure Rate} = \frac{F + A}{T},
\]
where \(F\) is the number of frames with detection failure, \(A\) is the number of extra hands falsely detected and \(T\) is the total number of frames.

\subsection{Landmark Detection Accuracy}
Percentage of Correct Keypoints (PCK) \cite{chen2020monocular} is adopted. A keypoint is considered correctly detected if it lies within a threshold distance \(\alpha\) from the ground truth:
\[
\text{PCK}(\alpha) = \frac{1}{N} \sum_{i=1}^{N} I\!\left(\frac{\| \hat{L}_i - L_i \|_2}{s_i} < \alpha\right),
\]
where \(\hat{L}_i\) is the predicted landmark, \(L_i\) is the ground truth, \(s_i\) is a scale normalization factor (e.g., hand width), \(\alpha=0.02\) is a pre-defined tolerance, and \(N=21\) is the total number of joints.

The Mean Per Joint Position Error (MPJPE) \cite{zhang2021estimation} quantifies the average Euclidean distance between predicted and ground truth joint positions:
\[
\text{MPJPE} = \frac{1}{N} \sum_{i=1}^{N} \left\| \hat{J}_i - J_i \right\|_2,
\]
where \(\hat{J}_i\) and \(J_i\) are the predicted and true positions of the \(i\)-th joint, and \(N\) is the total number of joints (e.g., 21 for hand landmarks). MPJPE is sensitive to both small and large deviations, making it effective for detecting subtle pose degradation.

\subsection{Event Counting Error}
Discrepancy between true and detected events is measured by:
\[
\text{Action Error} = \frac{|N_\text{GT} - N_\text{Detected}|}{N_\text{GT}}.
\]
\(N_\text{GT}\) is the number of true events, and \(N_\text{Detected}\) is the number detected.
\subsection{Medical Evaluation Error}
Mean Absolute Error (MAE) quantifies alignment with clinician labels.
\[
\text{MAE} = \frac{1}{N} \sum_{i=1}^{N} |y_i - \hat{y}_i|,
\]
where \(y_i\) is the true symptom severity score and \(\hat{y}_i\) is the predicted score.

\subsection{Label Evaluation}

Label quality directly impacts the clinical reliability of automated remote assessments. In this study, label consistency and ambiguity are assessed by analyzing inter-rater variability and weak supervision confidence, with particular emphasis on the zero-class (symptom severity score = 0), which is critical for real-world deployment due to its higher prevalence and susceptibility to false positives.

\paragraph{1. Uncertainty in Weak Supervision Labels.}
Uncertainty in probabilistic labels \(\hat{p}(y_i)\) derived from weak supervision is further analyzed. The analysis is conditioned on class:
\[
H_c = \frac{1}{|\mathcal{D}_c|} \sum_{x_i \in \mathcal{D}_c} -\sum_j \hat{p}_j(y_i) \log \hat{p}_j(y_i),
\]
where \(H_c\) denotes the average entropy for consensus-labeled class \(c\).

\paragraph{2. Inter-rater Variability and Agreement with Weak Supervision Labels}
Since symptom severity scores are ordinal, traditional accuracy-based metrics do not adequately reflect clinically meaningful differences between label predictions. For example, confusing a score of 0 with 1 is far less severe than confusing 0 with 4. To address this issue, we use Quadratic Weighted Kappa (QWK)~\cite{Cohen68}, which explicitly accounts for the magnitude of disagreement between raters or labeling methods based on ordinal distances. Formally, QWK is defined as:
\[
\kappa = 1 - \frac{\sum_{i,j} W_{ij} O_{ij}}{\sum_{i,j} W_{ij} E_{ij}},
\]
where \(O\) is the observed confusion matrix, \(E\) is the expected confusion matrix under random labeling, and \(W\) is the quadratic penalty matrix defined as:
\[
W_{ij} = \frac{(i - j)^2}{(K - 1)^2},
\]
with \(K = 5\) representing the number of classes. QWK scores range from 0 (chance agreement) to 1 (perfect agreement). In our analysis, we use QWK in two ways: first, to quantify inter-rater variability among clinical annotators, highlighting areas of greatest disagreement; second, to assess agreement between the aggregated weak supervision labels and individual expert annotations, validating how effectively our approach captures expert consensus.

\begin{table}[b]
\caption{Comparison of landmark detection, event counter, and medical symptom assessment Across Domains}
\label{tab:unified_metrics}
\vskip 0.15in
\begin{center}
\begin{small}
\begin{sc}
\begin{tabular}{lccc}
\toprule
Metric & Clinical & Home & Gap \\
\midrule
Failure Rate (\%) & 3.32$(\pm 10.33)$ & 26.5$(\pm 19.03)$ & 23.18 \\
EC Error (\%) & 20.44$(\pm 25)$  & 36.79$(\pm 23)$ & 16.35 \\
ME Error (MAE) & 0.69 & 0.83 & 0.14 \\
\bottomrule
\end{tabular}
\end{sc}
\end{small}
\end{center}

\end{table}

\section{Experiments and Results}
\subsection{Domain Gap: Clinical vs. Home-Recorded Data}

Clinical and home-recorded video data are compared across three key metrics: landmark detection reliability, action event counting error, and medical evaluation performance. This analysis quantifies the domain gap that arises when models trained in controlled environments are deployed in noisy, home-recorded settings.

\begin{figure}[t]
\begin{center}
\includegraphics[width=\linewidth]{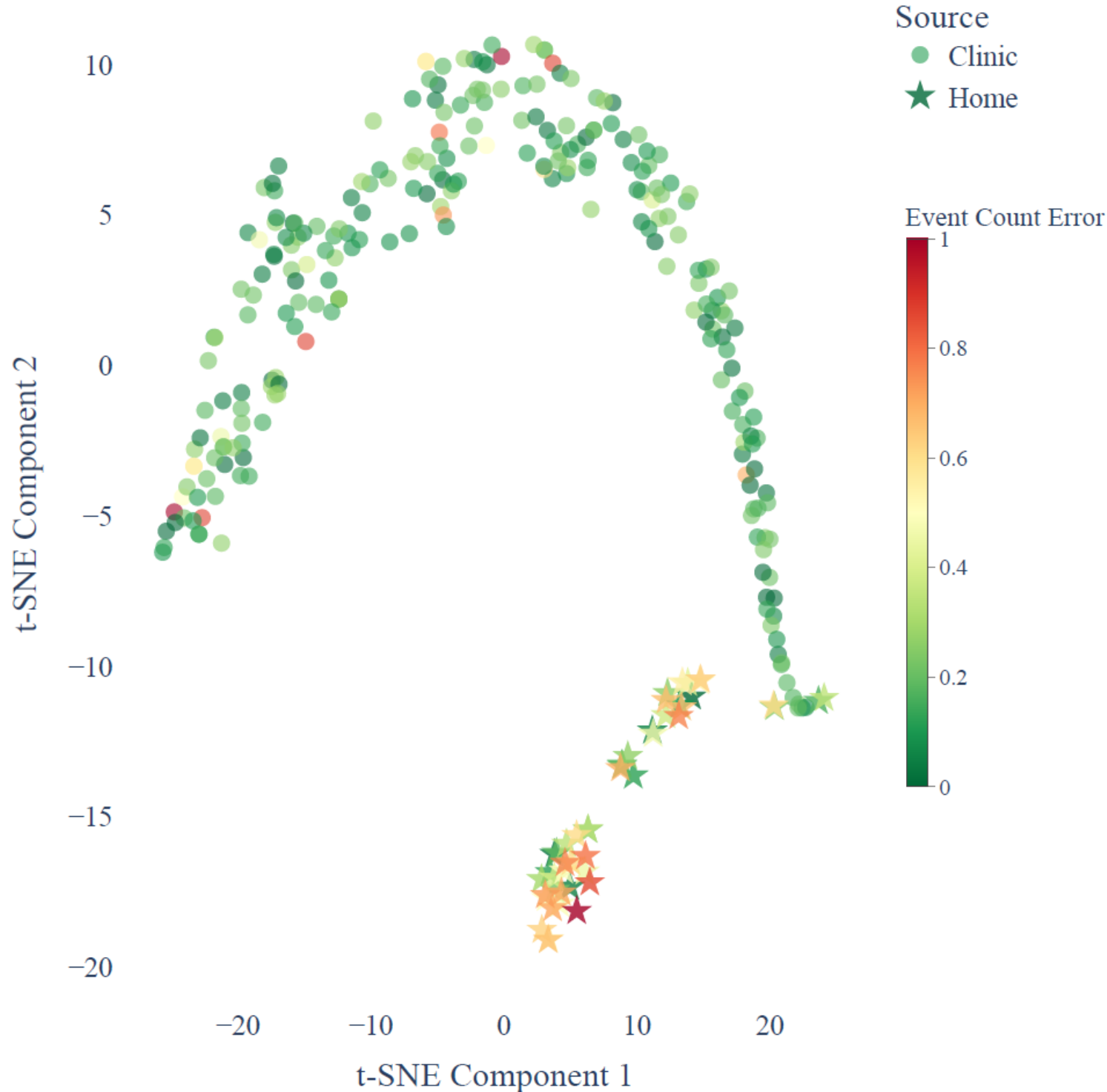}
\caption{t-SNE visualization of clinic vs home video features. Distinct clusters confirm distributional shift between domains.}
\label{fig:tsne}
\end{center}
\vskip -0.1in
\end{figure}

Figure~\ref{fig:tsne} highlights the t-SNE projection of video samples using quality-related features such as lighting conditions, frame rate, and contrast. These features were used to train the t-SNE model, revealing a clear domain shift between clinical (circles) and home-recorded (stars) data. The visualization shows distinct clusters, confirming substantial divergence in feature distributions across recording settings. Samples closer to the clinical cluster generally exhibit lower event counting error (ECE), suggesting that video quality features strongly influence downstream performance.
This distributional discrepancy has critical implications for remote health monitoring: models trained solely on well-controlled clinical data tend to underperform in home environments, where uncontrolled variables such as lighting, background clutter, and user behavior vary significantly. As shown in Table~\ref{tab:unified_metrics}, the home domain exhibits increased landmark detection failure rates and higher event count errors. Without domain adaptation or video quality control, such shifts can degrade model reliability and impair symptom evaluation. These findings emphasize the need for domain-aware pipelines that incorporate adaptive thresholds, quality filtering, or fine-tuning to maintain performance in real-world, at-home assessments.

\subsection{Degradation Study}

To assess robustness under real-world conditions, systematic degradations were applied to video quality parameters, including resolution, frame rate, brightness, contrast, scale, and noise. These manipulations were selected based on empirical artifacts observed in home recordings. A detailed summary of the applied degradations is provided in Appendix Table ~\ref{tab:video_manipulations}.

Table~\ref{tab:summary_degradations} summarizes the impact of each manipulation on MPJPE, PCK, LDFR, EC error, and statistical significance. The most substantial degradation was observed under low FPS (10), which caused a sharp increase in EC Err. (0.73) and a notable drop in PCK (0.61), confirming the sensitivity to temporal resolution.

In contrast, the resolution reduction (240p) maintained a high PCK (0.90) and a low EC error (0.16), suggesting that spatial downsampling alone does not significantly impair temporal event recognition. Interestingly, EC error. slightly decreased under some mild degradations, such as color distortion or Gaussian blur. These cases may introduce implicit smoothing or contrast enhancements that reduce spurious landmark flickering, leading to more stable event detection.
However, when the subject occupies a reduced portion of the frame (Minimize ×3), performance degrades markedly (LDFR: 0.28; EC Err.: 0.45), underscoring the importance of the subject's frame over absolute resolution.
Collectively, these findings offer actionable thresholds for automated quality control and guide practical recommendations for video capture in home settings.

\begin{table*}[h]
\caption{Summary of Most Impactful Degradations on Video Quality}
\label{tab:summary_degradations}
\vskip 0.15in
\begin{center}
\begin{small}
\begin{sc}
\begin{tabular}{lcccccc}
\toprule
Manipulation & MPJPE & PCK & LDFR & EC Err. & $p$-value \\
\midrule
Original Video & 0.00 (±0.00) & 1.00 (±0.00) & 0.03 (±0.12) & 0.20 (±0.23) & - \\
Frame Drops: 0.5 & \textbf{0.03 (±0.02)} & 0.43 (±0.20) & 0.00 (±0.02) & 0.34 (±0.20) & 0.00 \\
Resolution: 240p & 0.01 (±0.00) & 0.90 (±0.14) & 0.00 (±0.00) & 0.16 (±0.22) & 0.60 \\
Frame Rate: 10 fps & 0.02 (±0.02) & 0.61 (±0.23) & 0.01 (±0.05) & \textbf{0.73 (±0.13)} & 0.00 \\
Minimize: \texttimes3 & - & - & \textbf{0.28 (±0.39)} & 0.45 (±0.40) & 0.00 \\
Maximize: 1.5\texttimes & - & - & 0.03 (±0.12) & 0.32 (±0.36) & 0.00 \\
Gaussian Blur: 13 & 0.00 (±0.01) & \textbf{0.94 (±0.13)} & 0.00 (±0.03) & 0.15 (±0.23) & 0.03 \\
Brightness: -50 & 0.01 (±0.05) & 0.90 (±0.21) & 0.03 (±0.12) & 0.17 (±0.22) & 0.11 \\
Contrast: \texttimes1.5 & 0.01 (±0.01) & 0.86 (±0.18) & 0.01 (±0.03) & 0.25 (±0.32) & 0.13 \\
Color Distortion: -15 & 0.01 (±0.01) & \textbf{0.94 (±0.13)} & 0.03 (±0.12) & 0.25 (±0.32) & 0.01 \\
\bottomrule
\end{tabular}
\end{sc}
\end{small}
\end{center}
\vskip -0.1in
\end{table*}

\begin{table}[h]
\caption{Comparison of Model Performance Across Labeling Strategies. }
\label{tab:label_model_performance}
\vskip 0.15in
\begin{center}
\begin{small}
\begin{sc}
\begin{tabular}{l|cccc}
\toprule
Label Source & MAE (\textdownarrow) & F1 (Std) & Acc. & $FPR_0$ \\
\midrule
Single Rater     & 0.91 & 0.26 (0.12) & 0.30 & 0.13 \\
Majority Vote    & 0.82 & 0.43 (0.07) & 0.48 & 0.10 \\
HiLWS Raters     & 0.46 & 0.59 (0.04) & \textbf{0.61} & 0.08 \\
HiLWS Full       & \textbf{0.42} & \textbf{0.60} (0.04) & \textbf{0.61} & \textbf{0.06} \\
\bottomrule
\end{tabular}
\end{sc}
\end{small}
\end{center}
\vskip -0.1in
\end{table}
\subsection{Sensitivity Analysis}

A sensitivity analysis was conducted to determine practical thresholds for video parameters. The focus was on identifying the minimum requirements for accurate pose estimation and clinical assessment in home-recorded settings.

\paragraph{1. Quality Threshold Determination.}
Videos were systematically degraded along two key dimensions: frame rate and relative hand area. These manipulations enabled identification of critical thresholds beyond which pose estimation accuracy and event count reliability deteriorated. As shown in detail in Appendix Figure ~\ref{fig:appendix_threshold}, there is a clear decline in event count and landmark detection performance below specific thresholds. 

The analysis revealed that: (1) \textit{Frame rates below 24 FPS} significantly increased MPJPE and event counting error.
(2) The optimal relative \textit{hand area lies between 10\% and 35\% of the frame area.} Ratios below 5\% or above 40\% increase LDFR and EC Error.

These thresholds define the minimum technical standards for reliable home assessments. Videos below these standards showed higher errors in clinical metrics. The findings support enforcing video quality guidelines and using real-time feedback (e.g., frame rate or visibility alerts) to improve data usability and diagnostic accuracy.

\paragraph{2. Demographic Visual Fairness.}
No consistent landmark mismatch attributable to participant gender was observed. However, certain visual attributes associated with age and video conditions impacted detection reliability. Specifically, some elderly individuals with deep hand wrinkles were erroneously detected as having two hands due to false contour boundaries. Additionally, although skin tone itself did not directly impair landmark detection, failures were more frequent when sharpness was low, the hand was out of focus, or the background color closely resembled skin tone—leading to poor separation between hand and background. These findings suggest that landmark extraction models are sensitive to compounded visual artifacts rather than demographic attributes in isolation.

\subsection{Ablation Studies}
We discuss how to systematically ablate components of the video assessment pipeline to quantify their contribution to final performance.

\paragraph{Landmark Detection Confidence Threshold.}
\begin{figure}[h]
\centering
\includegraphics[width=0.8\linewidth]{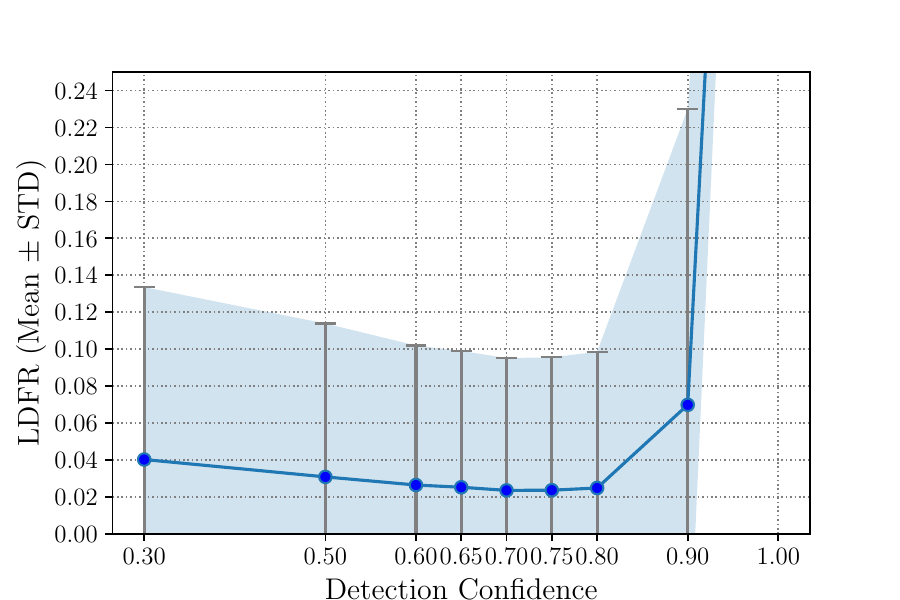}
\caption{Failure rate (mean ± std) for hand landmark detection across varying confidence thresholds. Confidence thresholds between 0.7 and 0.8 minimize LDFR.}
\label{fig:confidence_plot}
\end{figure}
Hand landmark detection models such as MediaPipe rely on internal confidence thresholds to determine the validity of a detected pose. These thresholds significantly influence downstream tasks such as motion tracking and tapping count. Parameter sweeps across a range of confidence values were performed to assess their effect on detection reliability and failure rate.

LDFR was defined as the proportion of frames in which the system either failed to detect the specified hand or incorrectly detected multiple instances of the same hand (e.g., two left hands). Each video was processed for a target hand (left or right), and error statistics were aggregated across clinical videos. The resulting metrics represent the mean and standard deviation of error rates across the dataset.

The results in Figure~\ref{fig:confidence_plot} suggest that a confidence threshold between 0.7 and 0.8 provides the optimal trade-off between detection robustness and strictness. Very high thresholds (e.g., 1.0) lead to excessive rejection of valid frames, while very low values (e.g., 0.3) yield noisy and unreliable detections.

\paragraph{HiLWS Labeling Strategy.}
Table~\ref{tab:label_model_performance} presents a comparative evaluation of the performance of the prediction of the severity of the symptom using four labeling strategies: single expert rater, majority vote, \textbf{HiLWS} with fusion of the rater only and the full \textbf{HiLWS} system that integrates both rater consensus and machine learning guidance. The evaluation was conducted on a held-out gold-standard balanced test set comprising 60 individuals whose labels exhibited complete agreement across all annotators. These labels were manually validated to ensure correctness and were not used in any part of the training process. This test set provides a high-confidence benchmark for assessing model performance.

The full \textbf{HiLWS} system achieves the lowest MAE, highest F1 score, and lowest false positive rate for the "no symptom" class ($FPR_0$), indicating superior alignment with expert annotations and robustness in low-symptom cases. These results highlight that Full \textbf{HiLWS} produces more reliable and clinically aligned labels compared to other approaches.

\begin{figure}[h]
\caption{Quadratic Weighted Kappa (QWK) agreement matrix between five expert annotators and the \textbf{HiLWS} system.}
\vskip 0.2in
\begin{center}
\includegraphics[width=\linewidth]{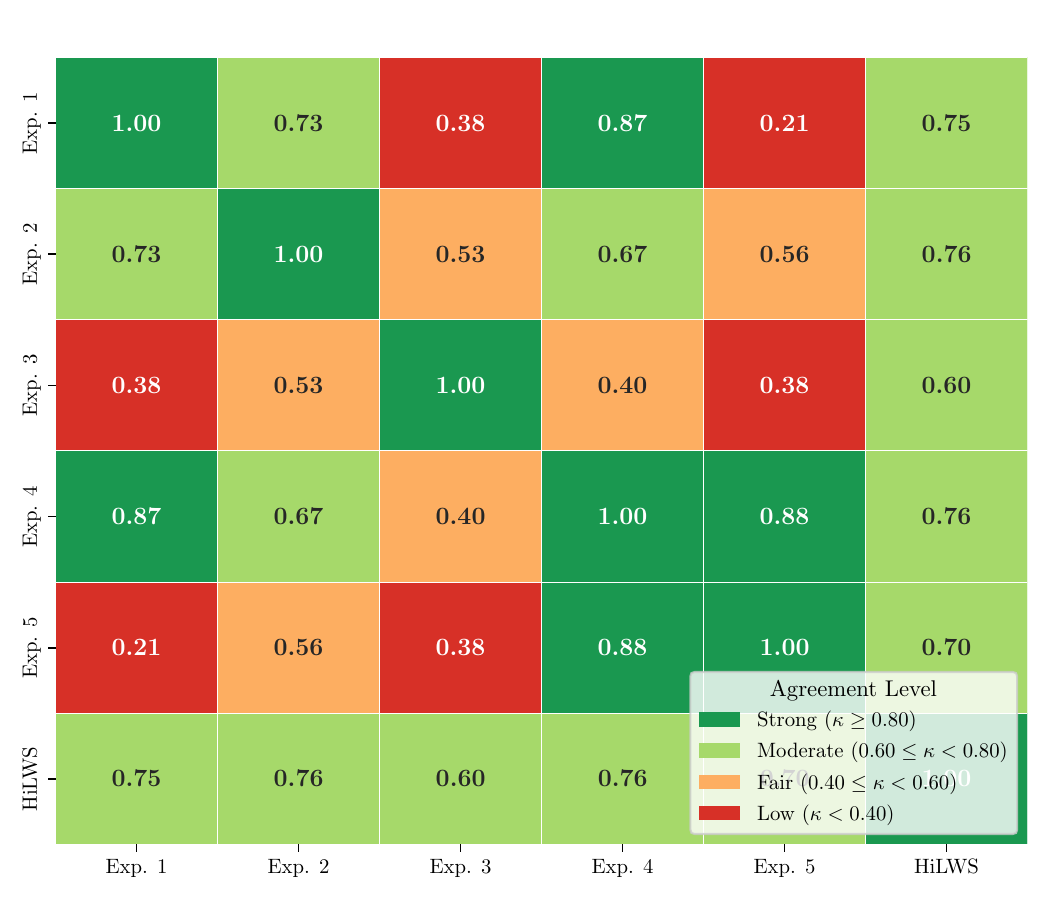}
\label{fig:qwk}
\end{center}
\vskip -0.2in
\end{figure}

\paragraph{Inter-Rater Agreement and HiLWS Alignment.}
Figure~\ref{fig:qwk} presents the pairwise Quadratic Weighted Kappa (QWK) scores between five expert raters and the \textbf{HiLWS} labeling system. \textbf{HiLWS} achieves moderate to strong agreement with all individual experts, with QWK scores ranging from 0.60 to 0.76. Notably, \textbf{HiLWS} aligns most closely with Expert~2 and Expert~4 (\(\kappa=0.76\)), suggesting it captures consensus patterns present in more consistent annotators. Among experts, the highest agreement is observed between Expert~4 and Expert~5 (\(\kappa=0.88\)), while the lowest is between Expert~1 and Expert~5 (\(\kappa=0.21\)), indicating variability in rater interpretation. Overall, the results demonstrate that \textbf{HiLWS} provides a stable approximation of expert-level labels, mitigating inter-rater disagreement and improving annotation reliability.

\paragraph{Inter-Rater Variability and Label Ambiguity.}
Table~\ref{tab:hilws_quality} reports the mean entropy of \textbf{HiLWS} generated probabilistic labels for each class.

The practical significance of these results lies in the clear reduction in label entropy when adding model-based fusion, particularly in symptom severity scores 0 and 1. These classes, typically associated with subtle or absent motor symptoms, are the most challenging to assess. By integrating machine learning into the labeling process, \textbf{HiLWS} effectively reduces ambiguity and sharpens the decision boundaries in these difficult cases. This shows that combining rater input with model guidance leads to more robust and confident labels, especially where human disagreement is highest. These insights directly support the design choice of hybrid labeling: ML helps resolve low-consensus cases and improves label quality in weakly supervised pipelines.

\begin{table}[h]
\caption{Mean entropy ($H_c$) across symptom severity scores for \textbf{HiLWS} and its No-ML variant. Lower values indicate reduced ambiguity in class assignments.}
\label{tab:hilws_quality}
\vskip 0.15in
\begin{center}
\begin{small}
\begin{sc}
\begin{tabular}{l|ccccc}
\toprule
\textbf{Symptom Severity} & \textbf{0} & \textbf{1} & \textbf{2} & \textbf{3} & \textbf{4} \\
\midrule
HiLWS (raters) $H_c$ & 0.28 & 0.57 & 0.50 & 0.38 & \textbf{0.16} \\
HiLWS (Full) $H_c$ & \textbf{0.18} & \textbf{0.34} & \textbf{0.36} & \textbf{0.20} & 0.29 \\
\bottomrule
\end{tabular}
\end{sc}
\end{small}
\end{center}
\vskip -0.1in
\end{table}

\section{Conclusion and Future Work}

This work introduces \textbf{HiLWS}, a cascaded human-in-the-loop weak supervision framework for curating structured video data collected in both clinical and home environments for remote neurological assessment. By addressing key challenges such as visual degradation and inter-rater disagreement, we demonstrate that carefully designed data curation strategies are essential for enabling robust machine learning in real-world healthcare applications.

Our modular pipeline incorporates adaptive video filtering, pose parameter tuning, task-aware segment selection, and probabilistic label aggregation. Through a series of degradation and sensitivity analyses, we identify domain-specific thresholds—such as minimum frame rate and hand visibility—that critically impact pose estimation quality and clinical prediction accuracy. The \textbf{HiLWS} framework improves upon majority voting by incorporating annotator reliability and label uncertainty, producing more robust labels particularly in ambiguous or low-symptom cases.

Beyond the specific use case of Parkinson's disease, this study highlights broader data-centric insights: the importance of context-aware evaluation metrics, the limitations of naive label fusion under domain shift, and the need for scalable human-in-the-loop strategies in high-stakes environments. These lessons are directly aligned with the goals of the DataWorld workshop, which offers guidance in developing generalizable data curation workflows across domains.

Future work will extend \textbf{HiLWS} to support federated learning across institutions and incorporate real-time quality feedback during video capture. Additionally, we aim to adapt this framework to other clinical contexts such as stroke rehabilitation, Tourette syndrome, and psychiatric motor side effects, demonstrating its broader utility across remote health monitoring scenarios.

\section*{Acknowledgements}

This research was supported by a Collaborative Health Research Project grant from the Canadian Institutes of Health Research (CIHR) in collaboration with the Social Sciences and Humanities Research Council of Canada (SSHRC) and the Natural Sciences and Engineering Research Council of Canada (NSERC) [Grant No. GR013210], as well as by the Pacific Parkinson’s Research Institute [Grant No. GR005879].
We gratefully acknowledge the participants of the Canadian Open Parkinson Network (C-OPN) registry for their valuable contributions.

\section*{Impact Statement}

This work presents a robust data curation and labeling framework for enabling remote medical assessments using home-recorded video data. By supporting the reliable analysis of structured motor tasks, such as those used in Parkinson's disease evaluations, outside of clinical settings, the proposed approach expands access to neurological care for individuals with limited mobility, including older adults, postsurgical patients, and those with chronic conditions that require ongoing monitoring.

The social impact lies in reducing the reliance on in-person evaluations, thus minimizing logistical barriers and healthcare disparities. However, variations in environmental conditions (e.g., lighting, background clutter) and individual characteristics (e.g., skin tone, hand morphology) can introduce algorithmic bias or reduce system reliability. To address these risks, the framework integrates quality-based video filtering, demographic sensitivity awareness, and a human-in-the-loop weak supervision process that promotes label robustness and ethical oversight.

Although not a replacement for clinical judgment, this system provides a scalable, privacy-preserving, and equitable tool to improve remote health surveillance. Its design reflects a broader commitment to trustworthy AI and contributes to the development of generalizable, context-aware data curation practices in real-world healthcare applications.

\bibliographystyle{icml2025}
\bibliography{main}
\newpage
\appendix
\onecolumn
\section{Dataset}
\renewcommand{\thetable}{A.\arabic{table}}
\setcounter{table}{0}
\begin{table}[h]
\centering
\caption{Demographic characteristics of participants recruited in the clinical setting.}
\label{tab:demography}
\begin{tabular}{lccc}
\toprule
Group &   N &  Mean Age ($\pm$ STD.) & Sex (M/F) \\
\midrule
   HC &  51 &      66.4 ($\pm$ 9.27) &     15/44 \\
   PD & 176 &      69.8 ($\pm$ 9.92) &    118/58 \\
Total & 227 &      69.0 ($\pm$ 9.5) &   133/102 \\
\bottomrule
\end{tabular}
\end{table}
\begin{table}[h]
\centering
\caption{Demographic characteristics of participants recruited at home.}
\label{tab:pd_unique_demo}
\begin{tabular}{lccc}
\toprule
Group &  N &  Mean Age ($\pm$ STD.) & Sex (M/F) \\
\midrule
   PD & 88 &      66.1 ($\pm$ 11.47) &     52/36 \\
\bottomrule
\end{tabular}
\end{table}

\newpage

\section{Video manipulations and their implementation parameters}

\renewcommand{\thetable}{B.\arabic{table}}
\setcounter{table}{0}

\begin{table}[h]
\caption{Summary of Video Manipulations and Implementation Details. This table outlines each degradation type applied to video frames for robustness evaluation. Implementations leverage OpenCV and FFmpeg, and simulate real-world video impairments encountered in remote assessments. Each method was implemented both frame-wise (Python/OpenCV) and stream-wise (FFmpeg) for flexibility.}
\label{tab:video_manipulations}
\vskip 0.15in
\begin{center}
\begin{small}
\begin{tabular}{p{3.2cm} | p{4.2cm} | p{3.8cm} | p{3.5cm}}
\toprule
\textbf{Manipulation} & \textbf{Technical Definition} & \textbf{Implementation} & \textbf{Parameters} \\
\midrule
Random Frame Drops & Stochastically discard frames to mimic transmission loss or sensor glitches. Frame count and fps are recomputed post drop. & Drop frames with probability $p$. & $p = 0.1$ to $0.5$ \\
\midrule
Resolution Reduction & Resize frames to simulate low-quality video acquisition and compression artifacts. & Downsample via `cv2.resize()` or `ffmpeg -vf scale`. & 1280×720 to 426×240 \\
\midrule
Frame Rate Reduction & Uniform subsampling of frames to lower effective FPS, simulating low temporal resolution. & Select every $k$-th frame based on target FPS. FFmpeg used with `-r`. & 60 to 10 FPS \\
\midrule
Minimize Subject & Simulate distant subject views by downscaling and padding. Maintains original frame size. & Resize then pad with black borders. & Scale = 1.2× to 3.0× \\
\midrule
Magnify Subject & Simulate proximity or cropped views by zooming into a centered or hand-centric ROI. & Crop center or detect hand via MediaPipe; enlarge and pad. & Zoom = 1.1× to 1.5× \\
\midrule
Add Gaussian Blur & Apply spatial smoothing with 2D Gaussian kernel to simulate defocus or motion blur. & Use `cv2.GaussianBlur()` or `ffmpeg -vf gblur`. & Kernel sizes: 3 to 9 \\
\midrule
Brightness Adjustment & Linearly shift intensity values to simulate lighting variation. & Use `cv2.convertScaleAbs()` or `ffmpeg -vf eq`. & $\pm$50 pixel units \\
\midrule
Contrast Adjustment & Apply affine scaling to pixel values to vary contrast. & Multiply by scalar $\alpha$ using OpenCV/FFmpeg. & $\alpha = 0.5$ to $2.0$ \\
\midrule
Color Distortion & Modify hue component in HSV color space to simulate illumination or white balance shifts. & Shift hue in `cv2.cvtColor()` and clip. & Hue $\in [-15, +15]$ \\
\bottomrule
\end{tabular}
\end{small}
\end{center}
\vskip -0.1in
\end{table}
\newpage
\section{Quality Threshold Determination}

\renewcommand{\thefigure}{C.\arabic{figure}}
\setcounter{figure}{0}
\begin{figure}[h]
\centering
\includegraphics[width=\textwidth]{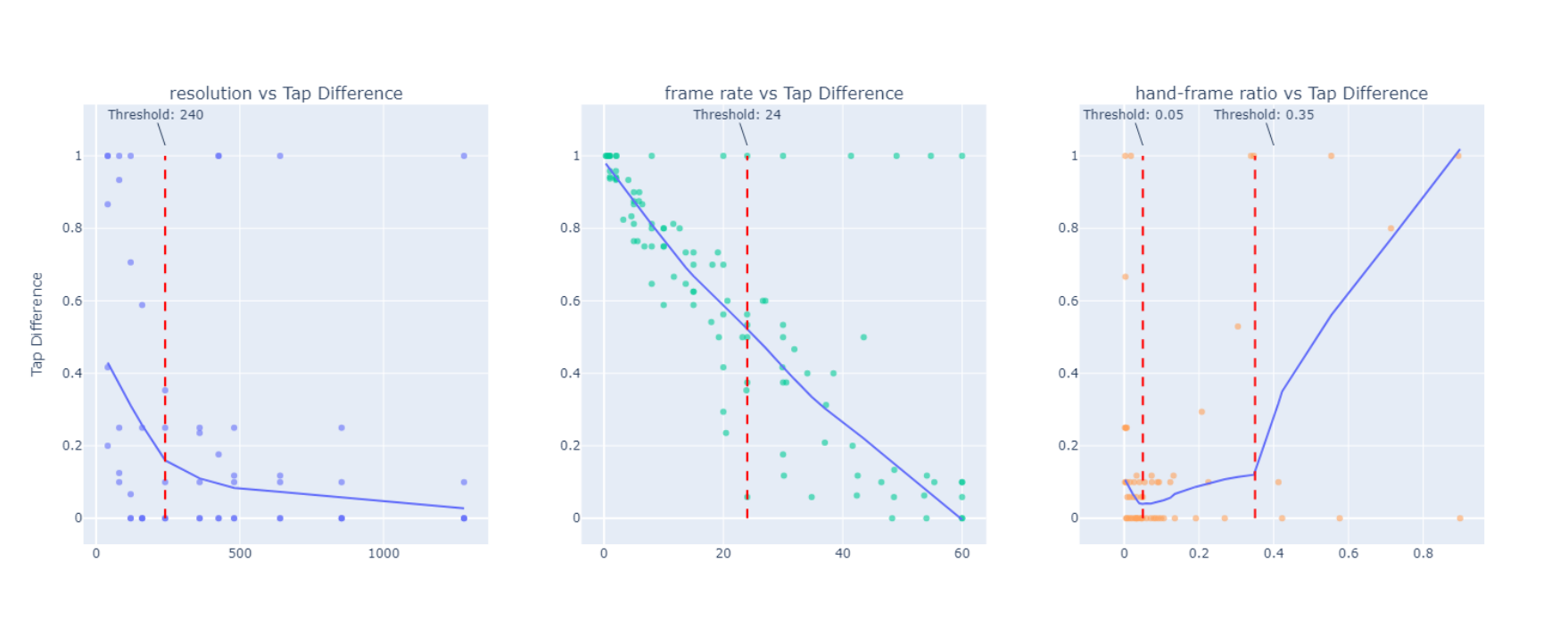}
\includegraphics[width=\textwidth]{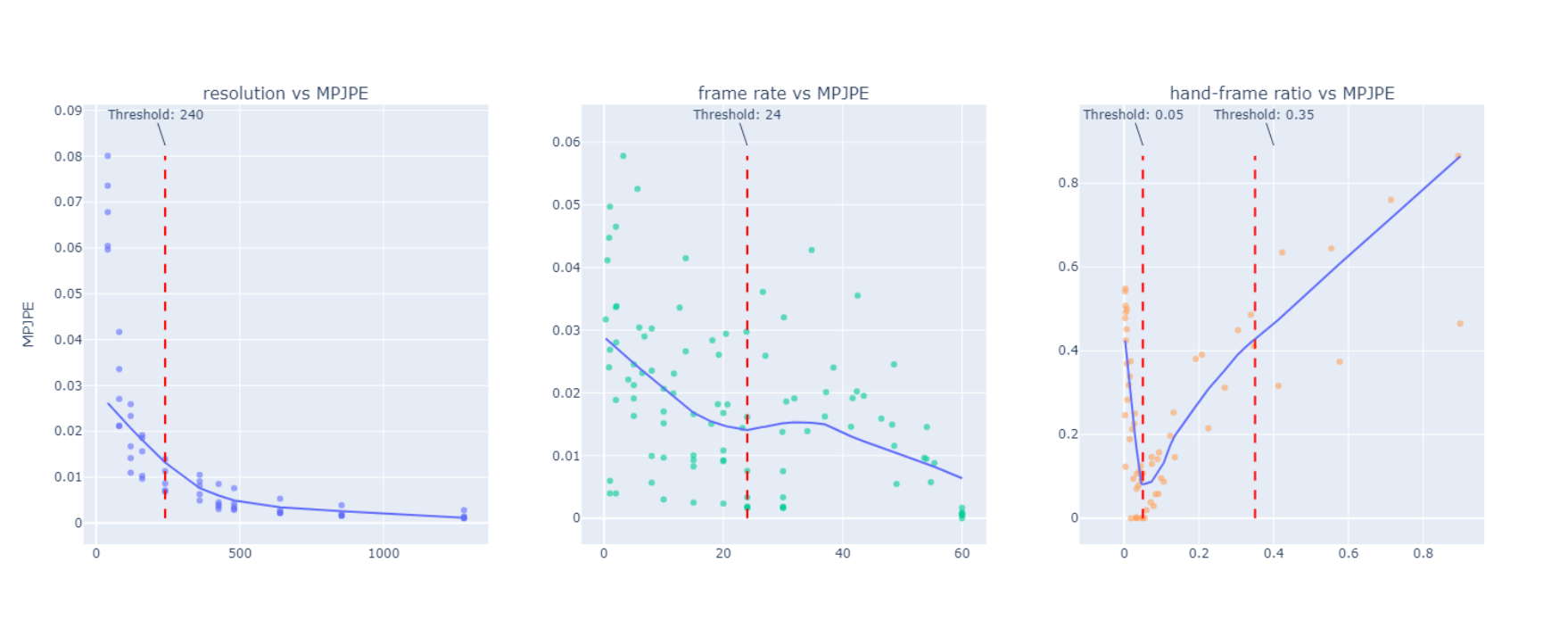}
\caption{\textit{Threshold detection analysis for video quality parameters and task performance.} The top row shows the effect of (left to right) resolution, frame rate, and relative hand area on event counting error. The bottom row presents their corresponding influence on landmark accuracy measured by Mean Per Joint Position Error (MPJPE). Red dashed lines mark the identified thresholds below which task and detection performance deteriorate significantly. In particular, frame rates below 24\,fps, resolutions below 240p, and relative hand area outside the [0.05, 0.35] range are associated with increased tap count errors and landmark inaccuracies.}
\label{fig:appendix_threshold}
\end{figure}

\end{document}